\documentclass[%
 reprint,
 amsmath,amssymb,
 aps,
 prl,superscriptaddress
]{revtex4-2}

\usepackage{graphicx}
\usepackage{dcolumn}
\usepackage{bm}
\usepackage[colorlinks=true,linkcolor=blue,citecolor=blue,urlcolor=blue]{hyperref}
\newcommand{\cur}{\mathrm{cur}}
\newcommand{\en}{\mathrm{en}}
\newcommand{\esc}{\mathrm{esc}}
\newcommand{\open}{\mathrm{open}}
\newcommand{\prop}{\mathrm{prop}}
\newcommand{\ev}{\mathrm{ev}}
\newcommand{\free}{\mathrm{free}}

\DeclareMathOperator{\sgn}{sgn}

\begin{document}

\title{Boundary-Robust Transmission Asymmetry as a Topological Signature in Open Floquet Lattices}

\author{Ren Zhang} \email{physiren@outlook.com}
 \affiliation{New Cornerstone Science Laboratory, Department of Physics,\\ \text{The Hong Kong University of Science and Technology,
Clear Water Bay, Kowloon 999077, Hong Kong, China}}

\author{Xiao-Yu Ouyang}
 \affiliation{New Cornerstone Science Laboratory, Department of Physics,\\ \text{The Hong Kong University of Science and Technology,
Clear Water Bay, Kowloon 999077, Hong Kong, China}}
 \affiliation{\text{Division of Chemistry and Chemical Engineering, California Institute of Technology, Pasadena, CA 91125, USA}}

\author{Xu-Dong Dai}
 \affiliation{New Cornerstone Science Laboratory, Department of Physics,\\ \text{The Hong Kong University of Science and Technology,
Clear Water Bay, Kowloon 999077, Hong Kong, China}}

\author{Xi Dai} \email{ daix@ust.hk }
\affiliation{New Cornerstone Science Laboratory, Department of Physics,\\ \text{The Hong Kong University of Science and Technology,
Clear Water Bay, Kowloon 999077, Hong Kong, China}}

\begin{abstract}
We identify a boundary-robust topological signature of open Floquet lattices: although nonadiabatic boundaries strongly reshape the transmission lineshape, the integrated left--right transmission asymmetry saturates to a plateau set by the bulk Floquet winding number. Its origin is a deep-bulk branch-population principle: in the long-sample limit, each propagating Floquet--Bloch branch is generically populated with unit weight, since true Floquet bound states are nongeneric. The robust observable is therefore the cumulative transmission imbalance rather than the boundary-sensitive transmission profile. We propose direct detection by cold-atom transmission spectroscopy. For electronic transport, the same asymmetry admits contact-model-dependent electrical readouts: a coherent Floquet--Landauer--B\"uttiker interpretation predicts a near-\(2ef\) response in weak SAW devices, whereas a blocking-factor post-processing yields a qualitatively different signal.
\end{abstract}

\maketitle

\textbf{\textit{Introduction.---}}
Periodically driven lattices support Floquet--Bloch bands and topological transport phenomena with no static analogue, including winding invariants and quantized pumping \cite{Shirley1965,Sambe1973,Thouless1983,Kitagawa2010,Rudner2013}. In realistic experiments, however, the driven region is probed through open geometries coupled to asymptotic leads. A central question is therefore whether bulk Floquet topology can be extracted directly from open-system transmission observables.

This is nontrivial because boundaries generically mix Floquet sidebands, while finite driven regions produce dense Fabry--P\'erot oscillations \cite{Born1999}. The relevant observable is thus not the pointwise transmission of a fixed sample, but the long-sample transmission obtained after local energy smoothing in the \(L\to\infty\) limit \cite{Bjoint}. In the benchmark case of an adiabatic boundary, the smoothed left- and right-incident transmission windows acquire a rigid offset whose width difference is \(C\hbar\omega\), with \(C\) the winding number.

Our main result is that this topological signature survives far beyond the adiabatic limit. Although nonadiabatic boundaries strongly distort the transmission lineshape, the integrated left--right transmission asymmetry remains nearly invariant and saturates to the same plateau. The robust quantity is therefore the cumulative asymmetry, not the detailed transmission profile.

We trace this robustness to a deep-bulk branch-population principle \cite{Bjoint}. In the long-sample limit, the smoothed transmission equals the population of propagating Floquet--Bloch branches deep inside the lattice. This population equals the escape probability of a wave packet initialized on the same branch at the sample center. Since true Floquet bound states require nongeneric matching conditions \cite{Yajima1983,Kaneta1987,Della2014,Bjoint}, each propagating branch is generically populated with unit weight. When an incident energy window selects an isolated Floquet band, the integrated asymmetry counts its net chirality and approaches \(C\hbar\omega\), independent of microscopic boundary details.

This suggests direct cold-atom transmission spectroscopy. For electronic transport, converting the same asymmetry into a zero-bias dc signal requires an additional contact prescription. Within a coherent Floquet--Landauer--B\"uttiker interpretation it predicts a near-\(2ef\) response in weak SAW devices \cite{Moskalets2002,Landauer1957,Buttiker1985,Datta1995}, whereas a blocking-factor post-processing yields a qualitatively different signal \cite{Wagner2000,Kim2003}. The robust result of this work is therefore the transmission asymmetry itself, while its electrical manifestation depends on contact modeling.

\textbf{\textit{Boundary-Robust Topological Spectral Asymmetry.---}}
We consider an open Floquet lattice of length \(L\), whose most direct observable is the transmission measured for left- and right-incident waves in asymptotically free leads. Throughout, we adopt dimensionless units with particle mass \(m=1/2\), \(\hbar=1\), and lattice period \(d=1\), yielding a recoil energy \(E_R=\hbar^2\pi^2/(2md^2)=\pi^2\). Let \(\mathcal T^{L}(E;L)\) and \(\mathcal T^{R}(E;L)\) denote the total transmission probabilities at incident energy \(E\). In extended driven samples, these pointwise spectra are dominated by dense Fabry--P\'erot oscillations \cite{Born1999} and exhibit strong sensitivity to \(L\), precluding their use as stable long-sample observables.
\begin{figure}[ht]
    \centering
    \includegraphics[width=1\linewidth]{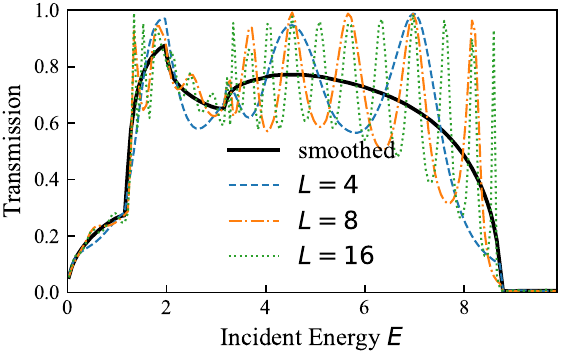}
    \caption{\textbf{Construction of the smoothed transmission spectrum.} Raw transmission curves for sample lengths \(L=4, 8, 16\) exhibit dense Fabry--P\'erot oscillations, whereas the energy-window smoothed curves rapidly converge. In the \(L\to\infty\) limit, the smoothed transmission becomes a well-defined local coarse-grained observable that effectively filters out finite-size interference.}
    \label{fig:FP_smooth}
\end{figure}
To extract a stable spectral signature, we coarse-grain the transmission over an energy window \(W_L(E;E_0)\) centered at \(E_0\), normalized as
\begin{equation}
\int_{\mathbb R} dE\, |W_L(E;E_0)|^2 = 1,
\label{eq:spec_window_norm}
\end{equation}
with width \(\Delta E(L)\) satisfying the asymptotic scaling
\begin{equation}
\Delta E(L)\to 0, \quad L\,\Delta E(L)\to\infty \quad \text{as } L\to\infty.
\label{eq:spec_window_scaling}
\end{equation}
The smoothed long-sample transmittance is then defined by
\begin{equation}
\bar{\mathcal T}^{L/R}(E_0)
=
\lim_{L\to\infty}
\int_{\mathbb R} dE\,
|W_L(E;E_0)|^2\,
\mathcal T^{L/R}(E;L).
\label{eq:spec_avg_trans}
\end{equation}
This construction filters out finite-size interference and defines the stable spectral observable used throughout.

Beyond finite-size effects, the transmission spectrum is highly sensitive to boundary details. We first consider a spatially adiabatic boundary that interpolates smoothly between free space and the Floquet lattice, thereby suppressing sideband mixing and backscattering \cite{Bjoint}. In this ideal limit, the smoothed spectra \(\bar{\mathcal T}^{L}(E)\) and \(\bar{\mathcal T}^{R}(E)\) form plateau-like windows whose widths correspond to the quasienergy bandwidths occupied by forward- (\(v_g>0\)) and backward- (\(v_g<0\)) propagating modes. Their width difference is strictly \(C\hbar\omega\), where \(C\) is the winding number of the Floquet band. This adiabatic limit thus establishes the cleanest spectroscopic benchmark for open Floquet topology.

\begin{figure}[ht]
    \centering
    \includegraphics[width=\columnwidth]{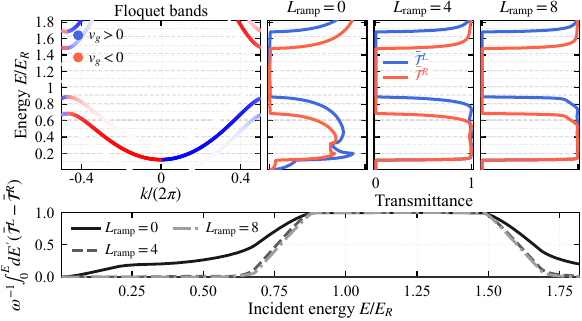}
    \caption{
    \textbf{Boundary robustness of the integrated spectral asymmetry.}
    For an open Floquet lattice with \(V(x,t)=\xi(x)V_{\rm bulk}(x,t)\) and \(V_{\rm bulk}(x,t)=0.8E_R\cos(2\pi x-\omega t)+0.2E_R\), the boundary envelope \(\xi(x)\) implements a linear ramp from 0 to 1 over a length \(L_{\rm ramp}\). While varying \(L_{\rm ramp}\) strongly distorts the smoothed transmission spectra \(\bar{\mathcal T}^{L/R}(E)\), the integrated asymmetry \(\mathbf A(E_0)\) remains invariant and saturates to a stable plateau. This plateau height equals the rigid spectral shift dictated by the bulk winding number in the adiabatic limit, demonstrating that the topological signature is immune to microscopic boundary details.}
    \label{fig:spectral_asymmetry}
\end{figure}

Our central result demonstrates that this topological signature remains robust well beyond the adiabatic regime. When boundaries deviate from adiabaticity, arbitrary nonadiabatic profiles strongly distort the fine structure of \(\bar{\mathcal T}^{L}(E)\) and \(\bar{\mathcal T}^{R}(E)\), yet the integrated left--right spectral asymmetry,
\begin{equation}
\mathbf A(E_0)
=
\int_0^{E_0} dE\,
\Bigl[
\bar{\mathcal T}^{L}(E)-\bar{\mathcal T}^{R}(E)
\Bigr],
\label{eq:area_asymmetry}
\end{equation}
A positive value of \(\mathbf A(E_0)\) means that the cumulative left-incident transmission exceeds the cumulative right-incident transmission over the interval \([0,E_0]\).
The integrated asymmetry remains invariant and saturates to a stable plateau at \(C\hbar\omega\) over a broad energy range. As shown in Fig.~\ref{fig:spectral_asymmetry}, while nonadiabatic boundaries reshape the transmittance windows, the cumulative asymmetry \(\mathbf A(E_0)\) remains invariant. The physically robust quantity is therefore not the transmission lineshape itself, but the integrated spectral imbalance. The adiabatic boundary merely resolves this invariant in its simplest form; the plateau persists even under strong boundary scattering.

This robustness demonstrates that the topological invariant is not encoded in boundary-sensitive transmission profiles. Its origin instead lies in the deep-bulk population of propagating Floquet--Bloch branches, as derived in the following section.

\textbf{\textit{Deep-Bulk Branch-Population Principle.---}}
To connect transmission observables to bulk topology, we adopt the quasienergy $\epsilon$ as the fundamental variable. The incident channel index $\alpha$ simultaneously encodes the laboratory energy and lead position (left/right), satisfying $E = \epsilon + n\omega$ ($n\in\mathbb{Z}$), and labels the incident state as $(\alpha,\epsilon)$. Propagating modes inside the lattice are labeled $(\mu,\epsilon)$, where $\mu$ distinguishes Floquet-Bloch branches, with quasienergy periodicity $(\mu,\epsilon+n\omega)\equiv(\mu,\epsilon)$.

The scattering state $\Psi_{\alpha}^{\mathrm{cur}}(x,t;\epsilon;L)$ with unit incident current from channel $\alpha$ admits the following asymptotic deep-bulk decomposition into a superposition of propagating branches (Floquet-Bloch waves) and boundary-localized evanescent modes:
\begin{equation}
\begin{split}
\Psi_{\alpha}^{\mathrm{cur}}(x,t;\epsilon;L)
&= \sum_{\mu\in \mathrm{prop}(\epsilon)}
d_{\mu\alpha}(\epsilon;L)\,
\Upsilon_{\mu}^{\mathrm{cur}}(x,t;\epsilon) \\
&\quad + \eta_\alpha(x,t;\epsilon;L),
\end{split}
\label{eq:deep_bulk_expansion}
\end{equation}
where $\Upsilon_{\mu}^{\mathrm{cur}}$ are current-normalized bulk Floquet-Bloch branches, and $\eta_\alpha$ contains evanescent waves localized near the boundaries that decay exponentially into the deep bulk. The coefficient $d_{\mu\alpha}$ directly quantifies the excitation strength of bulk branch $\mu$ by incident channel $\alpha$. To filter out finite-size interference, we define the smoothed branch population following the scheme in Eq.~\eqref{eq:spec_window_norm}:
\begin{equation}
p_{\mu\alpha}(\epsilon_0)
=
\lim_{L\to \infty}\int_{\mathbb R} d\epsilon\,
|W_L(\epsilon;\epsilon_0)|^2\,|d_{\mu\alpha}(\epsilon;L)|^2,
\end{equation}
and sum over all physically open incident channels ($\mathcal{A}_{\mathrm{open}} = \mathcal{A}^L_{\mathrm{open}} \cup \mathcal{A}^R_{\mathrm{open}}$) to obtain the total population on branch $\mu$:
\begin{equation}
p_{\mu}(\epsilon_0)
=
\sum_{\alpha\in\mathcal A_{\mathrm{open}}(\epsilon_0)}
p_{\mu\alpha}(\epsilon_0).
\label{eq:total_population}
\end{equation}

\textbf{Deep-bulk branch-population principle.}
The total population $p_\mu(\epsilon_0)$ exactly equals the escape probability $\mathcal{P}_{\mu}^{\mathrm{esc}}(\epsilon_0)$ of a wave packet prepared on branch $\mu$ deep inside the sample. This equivalence follows from the isomorphism between incoming and outgoing scattering channel spaces \cite{Howland1979,Howland2012}. Using $W_L$ with a suitable phase, one can construct a deep-bulk wave packet with spatial width $\ll L$. Its projection onto the open incoming channels converges to $p_\mu$, as the packet contains and probes only propagating Bloch components \cite{Bjoint,Supplemental}. Its projection onto the open outgoing channels yields the total escape probability. The channel-space isomorphism then directly implies
\begin{equation}
p_\mu(\epsilon_0)=\mathcal P_{\mu}^{\mathrm{esc}}(\epsilon_0).
\label{eq:branch_population_principle}
\end{equation}

In generic open geometries, permanent localization on a propagating branch (i.e., a true Floquet bound state) requires highly overdetermined boundary-matching conditions and thus occupies a set of measure zero in parameter space \cite{Bjoint,Supplemental,Yajima1983,Kaneta1987,Della2014}. Consequently, excited propagating branches are generically fully open, yielding an escape probability $\mathcal{P}_{\mu}^{\mathrm{esc}}$ that is generically unity. This leads to the generic result:
\begin{equation}
p_\mu(\epsilon_0) = 1,
\label{eq:generic_unity}
\end{equation}
which is independent of microscopic boundary details and directly underpins the boundary robustness of the transmission asymmetry.

This principle directly clarifies the topological origin of the cumulative spectral asymmetry. For a given channel $\alpha$, the smoothed transmission equals the sum over branch populations weighted by the sign of their group velocities. Exploiting the directional information encoded in $\alpha$, this can be written uniformly as:
\begin{equation}
\bar{\mathcal T}_\alpha(\epsilon) = \sum_{\mu} \sigma_\alpha \, p_{\mu\alpha}(\epsilon) \, \mathrm{sgn}(v_\mu),
\end{equation}
where $\sigma_\alpha = +1$ for left incidence ($\alpha \in \mathcal{A}^L_{\mathrm{open}}$) and $\sigma_\alpha = -1$ for right incidence ($\alpha \in \mathcal{A}^R_{\mathrm{open}}$), with $v_\mu = \partial \epsilon/\partial k$ the group velocity. 

When the target band is sufficiently isolated, choose an incident window \(\mathrm{Inc}\) such that
\[
\sum_{\alpha\in\mathrm{Inc}}p_{\mu\alpha}\approx
\begin{cases}
1,& \mu\in B_{\rm tar},\\
0,& \mu\notin B_{\rm tar}.
\end{cases}
\]
Within this window, the integrated left-right transmission asymmetry simplifies to:
\begin{equation}
\begin{split}
\mathbf{A}_{\mathrm{Inc}}
&= \int_{0}^{\hbar \omega} d\epsilon\, \sum_{\alpha \in \mathrm{Inc},\mu} \mathrm{sgn}(v_\mu)\, p_{\mu\alpha}(\epsilon) \\
&\approx \int_{0}^{\hbar \omega} d\epsilon\, \sum_{\mu\in B_{\mathrm{tar}}} \mathrm{sgn}(v_\mu)\,.
\end{split}
\label{eq:asymmetry_integral}
\end{equation}

The final summation strictly counts the net crossing number of the target band through a fixed quasienergy, which equals the bulk winding number $\mathcal{C}$. Therefore, $\mathbf{A}_{\mathrm{Inc}} \approx \hbar\omega\,\mathcal{C}$: the height of the spectral asymmetry plateau is directly locked to the bulk topological invariant. This mechanism establishes, from first principles, the topological nature of the boundary-robust transport signal in open Floquet lattices.

\textbf{\textit{Experimental Signatures.---}}
The transmission asymmetry can be probed directly in cold-atom spectroscopy and, under an additional contact prescription, converted into a zero-bias electrical signal in SAW devices.

\textbf{Cold-atom transmission spectroscopy.}
A quasi-1D atomic waveguide coupled to a traveling optical lattice directly realizes an open Floquet lattice with winding number one. The finite laser waist naturally provides a smooth spatial envelope, implementing an adiabatic boundary without artificial engineering \cite{Bloch2008,Schrader2001,Fabre2011}. This configuration directly reveals the plateau-like transmission windows and their width difference, as predicted in the adiabatic limit. Crucially, this setup allows flexible boundary shaping, enabling a direct test of the universal robustness of the integrated asymmetry \(\mathbf{A}(E_0)\) under nonadiabatic boundaries. The integrated observable is inherently insensitive to velocity spread, substantially relaxing monochromaticity requirements for the atomic source. Clear spectral resolution requires \(\delta E \ll \hbar\omega\), while maintaining \(\hbar\omega \lesssim V \lesssim E_R\) ensures a well-opened topological gap and cleanly separates transmission contributions from distinct photon sidebands.
\begin{figure}
    \centering
    \includegraphics[width=\columnwidth]{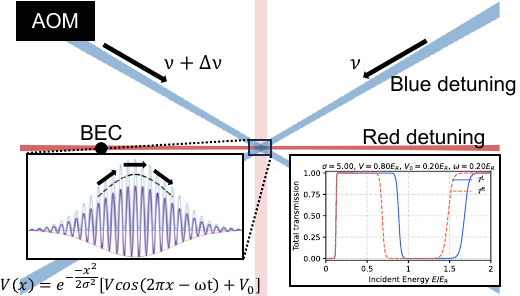}
    \caption{\textbf{Schematic of the cold-atom Bragg-scattering setup.} Two counter-propagating laser beams with a controlled frequency detuning, implemented via an acousto-optic modulator (AOM), generate a moving optical lattice. A Bose--Einstein condensate (BEC) serves as the scattering medium, confined by a quasi-1D red-detuned optical waveguide to realize a strictly one-dimensional scattering geometry.}
    \label{fig:cold_atom}
\end{figure}

\textbf{Electrical readout in SAW devices.}
\emph{Coherent scattering-state prescription.}
Assuming metallic leads attached to thermal reservoirs, the transmission asymmetry can be converted into a zero-bias current within a coherent Floquet--Landauer--B\"uttiker prescription \cite{Moskalets2002},
\begin{equation}
I_{\mathrm{coh}}
=
\frac{2e}{h}\int_0^{\infty} dE
\left[
\mathcal{T}^L(E) f_L(E)
-
\mathcal{T}^R(E) f_R(E)
\right],
\label{eq:Icoh}
\end{equation}
where \(\mathcal{T}^{L/R}(E)\) are the total current-normalized transmission probabilities for left- and right-incident scattering states. Here the energy origin is chosen at the bottom of the occupied one-dimensional subband, so \(\mu\) denotes the Fermi energy measured from that band bottom. At low temperature and zero bias, \(f_L=f_R\approx\Theta(\mu-E)\), so the integrated transmission asymmetry maps onto a near-\(2ef\) response when the Fermi level lies inside the transmission gap. A traveling SAW in a high-mobility nanowire on a piezoelectric substrate provides a possible realization of this weak-field extended-state regime, distinct from strong-field single-electron pumping dominated by localization and Coulomb blockade \cite{Fletcher2003,Shilton1995,Shilton1995-2,Shilton1996}.

\begin{figure}[ht]
    \centering
    \includegraphics[width=\columnwidth]{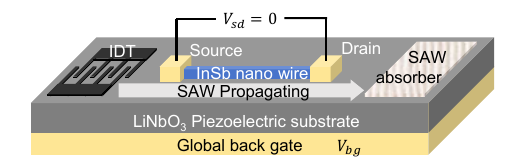}
    \caption{\textbf{Schematic of the SAW-driven zero-bias transport device.} An interdigitated transducer (IDT) on a piezoelectric substrate (e.g., LiNbO$_3$) launches a unidirectional surface acoustic wave, terminated by an absorber to suppress reflections. The traveling strain field induces a moving potential in a proximal nanowire (e.g., InSb). By tuning the chemical potential via a back-gate and maintaining zero source--drain bias, the resulting SAW-driven transport current is measured, providing a possible electrical readout of the integrated spectral asymmetry within the coherent-scattering interpretation.}
    \label{fig:SAW_device}
\end{figure}

\begin{figure}[ht]
    \centering
    \includegraphics[width=\columnwidth]{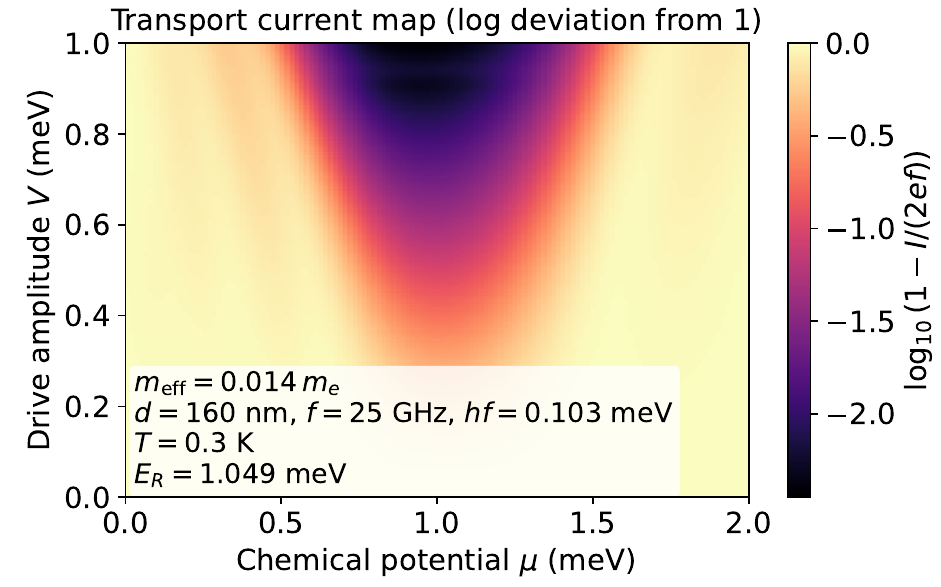}
    \caption{\textbf{Zero-bias current versus SAW drive strength and chemical potential within the coherent-scattering prescription.} Numerical results use hard-wall boundary conditions---a conservative estimate for which realistic soft boundaries should yield sharper quantization. A near-\(2ef\) response emerges over a broad parameter regime within this prescription, illustrating a possible electrical readout of the boundary-robust integrated transmission asymmetry.}
    \label{fig:SAW_transport}
\end{figure}

\emph{Alternative blocking-factor prescription.}
For comparison, one may post-process the same sideband-resolved transmission data within a phenomenological blocking-factor picture, in which the receiving contact supplies an explicit final-state occupation constraint. This prescription is not derived from the coherent scattering-state formalism above; rather, it represents an alternative contact model for the same scattering asymmetry \cite{Datta1995,Wagner2000,Kim2003,Oppenheimer1928,Seitz1969}.

Writing
\begin{equation}
\mathcal{T}^{L/R}(E)=\sum_{E_n>0}\mathcal{T}_n^{L/R}(E),
\qquad E_n=E+n\hbar\omega,
\label{eq:Tsideband}
\end{equation}
the corresponding current is
\begin{equation}
\begin{split}
I_{\mathrm{blk}} &= \frac{2e}{h} \int_0^{\infty} dE \sum_{E_n>0}
\biggl[
\mathcal{T}_n^L(E)\, f_L(E)\bigl(1-f_R(E_n)\bigr) \\
&\quad - \mathcal{T}_n^R(E)\, f_R(E)\bigl(1-f_L(E_n)\bigr)
\biggr].
\end{split}
\label{eq:Iblk}
\end{equation}
Here \(\mathcal{T}_n^{L}(E)\) [\(\mathcal{T}_n^{R}(E)\)] is the transmission probability for an electron incident at energy \(E\) from the left (right) lead to exit the opposite lead at energy \(E_n\).

The two prescriptions lead to qualitatively different zero-bias responses for the same sideband-resolved asymmetry: the coherent prescription yields a weakly temperature-dependent near-\(2ef\) plateau, whereas the blocking-factor prescription suppresses this plateau and produces a much stronger temperature dependence. We do not attempt to settle the microscopic validity of the two contact models here; rather, our point is that they lead to experimentally distinguishable electrical readouts of the same transmission asymmetry.

\begin{figure}[ht]
    \centering
    \includegraphics[width=\linewidth]{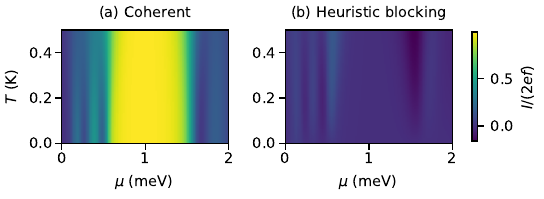}
    \caption{
    \textbf{Experimentally distinguishable transport predictions.}
    Left: zero-bias current from the coherent prescription [Eq.~(\ref{eq:Icoh})], showing a weakly temperature-dependent near-\(2ef\) plateau.
    Right: zero-bias current from the blocking-factor prescription [Eq.~(\ref{eq:Iblk})], obtained from the same sideband-resolved transmission data [Eq.~(\ref{eq:Tsideband})].
    The plateau is suppressed and the temperature dependence is much stronger.
    The driving amplitude is fixed at \(V=0.8E_R\), with all other parameters the same as in Fig.~\ref{fig:SAW_transport}.
    }
    \label{fig:transport_prescription_compare}
\end{figure}

\textbf{\textit{Discussion and conclusion.---}}
Our central result is that, beyond the adiabatic-boundary limit, the boundary-sensitive transmission profile gives way to a boundary-robust integrated transmission asymmetry whose plateau height is fixed solely by the winding number of the participating quasienergy band. The robust observable is therefore the cumulative left-right transmission imbalance, rather than the detailed transmission lineshape itself.

For cold atoms, this provides a direct transmission-spectroscopy signature of Floquet-band topology. For electrical transport, however, converting the same asymmetry into a measurable zero-bias current requires an additional contact prescription. Within a coherent Floquet--Landauer--B\"uttiker interpretation, the asymmetry maps onto a near-\(2ef\) response in a weak-field SAW setup, whereas an alternative blocking-factor prescription yields a qualitatively different signal. The robust result of this work is therefore the transmission asymmetry itself, while its electrical manifestation is contact-model dependent.

\section*{Acknowledgments}

X.D. Dai acknowledges the support from the New Cornerstone Foundation. X. Dai is supported by the New Cornerstone Foundation and a fellowship and a CRF award from the Research Grants Council of the Hong Kong Special Administrative Region, China (Projects No. HKUST SRFS2324-6S01 and No. C7037-22GF).

\textit{Data Availability.}---  The source code and numerical scripts that support the findings of this study are available on GitHub \cite{Github_data}.

\appendix

\bibliography{solve}

\clearpage

\setcounter{section}{0}
\setcounter{equation}{0}
\setcounter{figure}{0}
\setcounter{table}{0}

\renewcommand{\thesection}{S\arabic{section}}
\renewcommand{\thesubsection}{S\arabic{section}.\arabic{subsection}}
\renewcommand{\theequation}{S\arabic{equation}}
\renewcommand{\thefigure}{S\arabic{figure}}
\renewcommand{\thetable}{S\arabic{table}}

\twocolumngrid
\begin{center}
{\large\bf Supplemental Material}
\end{center}
\vspace{0.5em}

\section{Deep-bulk branch-population principle}
\label{sec:supp_deep_bulk}

This section proves the deep-bulk branch-population principle used in the main text.
The purpose is to show that the total long-sample population \(p_\mu(\epsilon_0)\) of a propagating Floquet--Bloch branch equals the escape probability of a wave packet prepared on that branch deep inside the sample:
\begin{equation}
p_\mu(\epsilon_0)=\mathcal P_\mu^{\esc}(\epsilon_0).
\label{eq:supp_branch_population_principle}
\end{equation}
The proof uses only the standard scattering-state structure of an open Floquet problem.
No specific transfer-matrix construction is needed.

\subsection{Branch-resolved scattering states}
\label{subsec:supp_branch_scattering}

We adopt the notation of the main text.
The quasienergy \(\epsilon\) is the fundamental spectral variable.
The incident channel index \(\alpha\) simultaneously encodes the incident lead and the Floquet sideband, so that the corresponding laboratory energy is
\begin{equation}
E=\epsilon+n\omega,
\qquad n\in\mathbb Z,
\label{eq:supp_E_alpha}
\end{equation}
with \(n\) included in \(\alpha\).
Thus \((\alpha,\epsilon)\) labels a physical incident state.
Propagating modes inside the lattice are labeled by \((\mu,\epsilon)\), where \(\mu\) distinguishes Floquet--Bloch branches.

Let
\begin{equation}
\Psi_{\alpha}^{\cur}(x,t;\epsilon;L)
\label{eq:supp_scatt_state_cur}
\end{equation}
be the scattering state generated by a unit incident current in channel \(\alpha\) for a driven segment of length \(L\).
In a regular quasienergy window around \(\epsilon_0\), the propagating Floquet--Bloch branches have nonzero group velocities
\[
v_\mu(\epsilon)=\frac{\partial\epsilon_\mu}{\partial k}.
\]
We choose the bulk branch states
\(\Upsilon_\mu^{\cur}(x,t;\epsilon)\) to be current-normalized, so that branch \(\mu\) carries unit current with sign
\begin{equation}
j_\mu=\sgn(v_\mu).
\label{eq:supp_branch_current_sign}
\end{equation}
Physically, this is the usual statement that flux equals probability per unit cell times group velocity, \(j_\mu=\rho_\mu v_\mu\), where \(\rho_\mu\) is the probability weight in one spatial period.
We use this relation only to identify the current-normalized convention with the quasienergy-normalized convention; its derivation is given in Ref.~\cite{Bjoint}.

Choose a point \(x_c(L)\) in the homogeneous part of the driven region such that both distances to the boundaries grow linearly with \(L\):
\begin{equation}
x_c(L)\to\infty,
\qquad
L-x_c(L)\to\infty.
\label{eq:supp_xc_deep}
\end{equation}
Near this point, every physical scattering state has the deep-bulk branch expansion
\begin{equation}
\begin{split}
\Psi_{\alpha}^{\cur}(x,t;\epsilon;L)
&=
\sum_{\mu\in\prop(\epsilon)}
d_{\mu\alpha}(\epsilon;L)\,
\Upsilon_{\mu}^{\cur}(x,t;\epsilon)
\\
&\quad
+\eta_\alpha(x,t;\epsilon;L).
\end{split}
\label{eq:supp_deep_bulk_expansion}
\end{equation}
Here \(\eta_\alpha\) consists of evanescent components localized near the boundaries.
Since \(x_c(L)\) lies deep in the sample, \(\eta_\alpha\) is exponentially small in the region probed below.
The coefficient \(d_{\mu\alpha}\) is the amplitude with which the incident channel \(\alpha\) populates the propagating bulk branch \(\mu\).

At fixed \(L\), the quantity \(|d_{\mu\alpha}(\epsilon;L)|^2\) still contains finite-size interference.
We therefore define branch populations using the same shrinking-window smoothing as in the main text.
Let \(W_L(\epsilon;\epsilon_0)\) be normalized by
\begin{equation}
\int_{\mathbb R}d\epsilon\,
|W_L(\epsilon;\epsilon_0)|^2=1,
\label{eq:supp_W_norm}
\end{equation}
with width \(\Delta\epsilon(L)\) satisfying
\begin{equation}
\Delta\epsilon(L)\to0,
\qquad
L\,\Delta\epsilon(L)\to\infty
\qquad
(L\to\infty).
\label{eq:supp_W_scaling}
\end{equation}
The channel-resolved branch population is
\begin{equation}
p_{\mu\alpha}(\epsilon_0)
=
\lim_{L\to\infty}
\int_{\mathbb R}d\epsilon\,
|W_L(\epsilon;\epsilon_0)|^2
|d_{\mu\alpha}(\epsilon;L)|^2 .
\label{eq:supp_pmualpha_def}
\end{equation}
Summing over all physically open incident channels gives the total population on branch \(\mu\):
\begin{equation}
p_\mu(\epsilon_0)
=
\sum_{\alpha\in\mathcal A_{\open}(\epsilon_0)}
p_{\mu\alpha}(\epsilon_0),
\label{eq:supp_pmu_def}
\end{equation}
where
\[
\mathcal A_{\open}
=
\mathcal A_{\open}^{L}
\cup
\mathcal A_{\open}^{R}.
\]

For the projection argument below, it is convenient to use quasienergy-normalized branch states.
We denote them by \(\Upsilon_\mu^{\en}(x,t;\epsilon)\), with
\begin{equation}
\langle\!\langle
\Upsilon_\mu^{\en}(\epsilon)
|
\Upsilon_\nu^{\en}(\epsilon')
\rangle\!\rangle
=
\delta_{\mu\nu}\delta(\epsilon-\epsilon').
\label{eq:supp_energy_norm}
\end{equation}
Here the double bracket means the Floquet inner product averaged over one drive period and integrated over space:
\begin{equation}
\langle\!\langle\psi|\phi\rangle\!\rangle
=
\frac{1}{\tau}\int_0^\tau dt\int_{\mathbb R}dx\,
\psi(x,t)^*\phi(x,t).
\label{eq:supp_double_bracket}
\end{equation}
The current-normalized and quasienergy-normalized conventions differ only by a fixed normalization factor on each regular branch.
With the convention of Ref.~\cite{Bjoint},
\begin{equation}
\Upsilon_\mu^{\cur}(\epsilon)
=
\sqrt{2\pi}\,
\Upsilon_\mu^{\en}(\epsilon).
\label{eq:supp_cur_en_relation}
\end{equation}
The same factor relates the unit-current and quasienergy-normalized scattering states, so the branch coefficients \(d_{\mu\alpha}\) are the same in either convention.

\subsection{Single-branch deep-bulk wave packet}
\label{subsec:supp_single_branch_packet}

We now prepare a wave packet on one propagating branch \(\mu\), centered at the deep-bulk point \(x_c(L)\).
Using the same envelope \(W_L(\epsilon;\epsilon_0)\), choose its phase so that the packet is spatially centered at \(x_c(L)\), and define
\begin{equation}
\Psi_{\mu,W}^{(L)}(x,t)
=
\int_{\mathbb R}d\epsilon\,
W_L(\epsilon;\epsilon_0)\,
\Upsilon_\mu^{\en}(x,t;\epsilon).
\label{eq:supp_single_branch_packet}
\end{equation}
Because the energy window shrinks while \(L\Delta\epsilon(L)\to\infty\), the packet width \(\ell(L)\) satisfies
\begin{equation}
\ell(L)\to\infty,
\qquad
\frac{\ell(L)}{L}\to0 .
\label{eq:supp_packet_width}
\end{equation}
Thus the packet is broad on microscopic scales, so it cleanly represents a branch wave packet, but remains asymptotically small compared with the full sample length.
It is therefore localized in the deep bulk and does not overlap the boundaries in the long-sample limit.

This scale separation is the key point.
In the spatial region where \(\Psi_{\mu,W}^{(L)}\) has support, the evanescent boundary pieces in Eq.~\eqref{eq:supp_deep_bulk_expansion} are exponentially small.
Therefore, when this packet is projected onto physical scattering states, the overlap is determined only by the propagating branch components.
Since the packet is constructed on branch \(\mu\), the regular-branch orthogonality in Eq.~\eqref{eq:supp_energy_norm} selects precisely the \(\mu\) component of the deep-bulk expansion.

\subsection{Projection proof of \(p_\mu=\mathcal P_\mu^{\esc}\)}
\label{subsec:supp_projection_proof}

Let \(\mathcal H_{\rm sc}\) be the scattering sector spanned by physical open-channel scattering states at the chosen quasienergy window.
The incoming and outgoing scattering states are two complete descriptions of the same sector.
Equivalently, the incoming and outgoing scattering channel spaces are isomorphic, as assumed in standard Floquet scattering theory \cite{Howland1979,Howland2012}.
Let \(P_{\rm sc}\) denote the orthogonal projection onto \(\mathcal H_{\rm sc}\).

We evaluate the scattering-sector weight of the packet in two ways:
\begin{equation}
\mathcal P_\mu^{\rm sc}(\epsilon_0)
=
\lim_{L\to\infty}
\langle\!\langle
\Psi_{\mu,W}^{(L)}
|
P_{\rm sc}
|
\Psi_{\mu,W}^{(L)}
\rangle\!\rangle .
\label{eq:supp_Psc_weight}
\end{equation}

First use the incoming scattering basis.
Let
\(\Psi_\alpha^{\en}(\epsilon;L)\) denote the quasienergy-normalized physical scattering state generated from incoming channel \(\alpha\).
Resolving the identity on the incoming open-channel basis gives
\begin{equation}
\begin{split}
&
\langle\!\langle
\Psi_{\mu,W}^{(L)}
|
P_{\rm sc}
|
\Psi_{\mu,W}^{(L)}
\rangle\!\rangle
\\
&\quad =
\sum_{\alpha\in\mathcal A_{\open}}
\int_{\mathbb R}d\epsilon\,
\left|
\langle\!\langle
\Psi_\alpha^{\en}(\epsilon;L)
|
\Psi_{\mu,W}^{(L)}
\rangle\!\rangle
\right|^2 .
\end{split}
\label{eq:supp_incoming_resolution}
\end{equation}
In the support of \(\Psi_{\mu,W}^{(L)}\), the scattering state has the deep-bulk propagating expansion
\begin{equation}
\Psi_\alpha^{\en}(x,t;\epsilon;L)
=
\sum_\nu
d_{\nu\alpha}(\epsilon;L)
\Upsilon_\nu^{\en}(x,t;\epsilon)
+
\text{exponentially small terms}.
\label{eq:supp_energy_deep_expansion}
\end{equation}
Using Eq.~\eqref{eq:supp_energy_norm}, only the \(\nu=\mu\) term overlaps with the packet in Eq.~\eqref{eq:supp_single_branch_packet}.
Therefore
\begin{equation}
\begin{split}
\mathcal P_\mu^{\rm sc}(\epsilon_0)
&=
\lim_{L\to\infty}
\sum_{\alpha\in\mathcal A_{\open}}
\int_{\mathbb R}d\epsilon\,
|W_L(\epsilon;\epsilon_0)|^2
|d_{\mu\alpha}(\epsilon;L)|^2
\\
&=
p_\mu(\epsilon_0).
\end{split}
\label{eq:supp_Psc_equals_pmu}
\end{equation}
Thus the projection of a single-branch deep-bulk packet onto the incoming scattering sector gives exactly the total branch population \(p_\mu\).

Now use the outgoing scattering basis instead.
Because incoming and outgoing scattering states span the same scattering sector, the projector \(P_{\rm sc}\) is unchanged.
However, in the outgoing representation, the same projection measures the part of the packet that eventually appears in the outgoing lead channels.
This is precisely the escape probability of the packet:
\begin{equation}
\mathcal P_\mu^{\rm sc}(\epsilon_0)
=
\mathcal P_\mu^{\esc}(\epsilon_0).
\label{eq:supp_Psc_equals_escape}
\end{equation}
Combining Eqs.~\eqref{eq:supp_Psc_equals_pmu} and
\eqref{eq:supp_Psc_equals_escape} proves
\begin{equation}
p_\mu(\epsilon_0)
=
\mathcal P_\mu^{\esc}(\epsilon_0).
\label{eq:supp_pmu_escape_final}
\end{equation}
This is the deep-bulk branch-population principle used in the main text.
\section{Generic openness from overdetermined bound-state matching}
\label{sec:supp_generic_openness}

We now explain why a propagating Floquet--Bloch branch is generically open.
By the branch-population principle proved in Sec.~\ref{sec:supp_deep_bulk},
\[
p_\mu(\epsilon_0)=\mathcal P_\mu^{\esc}(\epsilon_0).
\]
For a normalized packet prepared on branch \(\mu\), the only way to have
\(\mathcal P_\mu^{\esc}<1\) is for a nonzero part of the packet to remain trapped in a true bound sector.
Thus generic openness reduces to the absence of true Floquet bound trapping for propagating branches.

The point is simple.
A Floquet bound state must decay in both asymptotic leads.
However, once propagating lead channels are present, imposing decay on both sides requires more conditions than there are decaying amplitudes to tune.
This gives an overdetermined matching problem.
We formulate this statement explicitly below using only wave-function vectors and their linear spans.

\subsection{Floquet wave-function vector and free-space channels}
\label{subsec:supp_wfv_free_channels}

For fixed quasienergy \(\epsilon\), write the Floquet state as
\begin{equation}
\psi(x,t)
=
e^{-i\epsilon t}
\sum_{n\in\mathbb Z}
\varphi_n(x)e^{-in\omega t}.
\label{eq:supp_floquet_ansatz}
\end{equation}
The frequency components are collected into
\[
\bm\varphi(x)=(\ldots,\varphi_{-1}(x),\varphi_0(x),\varphi_1(x),\ldots)^T,
\]
and we define the Floquet wave-function vector
\begin{equation}
Y(x)
=
\begin{bmatrix}
\bm\varphi(x)\\
\bm\varphi'(x)
\end{bmatrix}.
\label{eq:supp_wfv_def}
\end{equation}
At fixed \(\epsilon\), the Schr\"odinger equation is a first-order spatial evolution equation for \(Y(x)\).
Therefore, specifying \(Y\) at one boundary uniquely determines the full solution everywhere in the sample.
If the left and right boundaries of the driven region are denoted by \(x_L\) and \(x_R\), we may write this spatial propagation schematically as
\begin{equation}
Y_R
=
F(\epsilon)Y_L,
\qquad
Y_L\equiv Y(x_L),\quad
Y_R\equiv Y(x_R),
\label{eq:supp_spatial_propagation}
\end{equation}
where \(F(\epsilon)\) is the finite propagation map across the driven region.
No further structure of \(F\) is needed for the rank argument below.

In the free leads, different Floquet sidebands decouple.
Let
\begin{equation}
E_n(\epsilon)=\epsilon+n\omega .
\label{eq:supp_sideband_energy}
\end{equation}
For \(E_n>0\), the \(n\)-th sideband is propagating, with
\[
k_n=\sqrt{E_n}.
\]
Using \(\mathbf e_n\) for the unit vector in frequency space, the right- and left-going free wave-function vectors may be written as
\begin{equation}
Y_n^{\rightarrow}
=
\begin{bmatrix}
\mathbf e_n\\
ik_n\mathbf e_n
\end{bmatrix},
\qquad
Y_n^{\leftarrow}
=
\begin{bmatrix}
\mathbf e_n\\
-ik_n\mathbf e_n
\end{bmatrix}.
\label{eq:supp_free_prop_channels}
\end{equation}
For \(E_n<0\), the sideband is evanescent, with
\[
\kappa_n=\sqrt{-E_n}.
\]
The two evanescent wave-function vectors are
\begin{equation}
Y_n^{+,\ev}
=
\begin{bmatrix}
\mathbf e_n\\
-\kappa_n\mathbf e_n
\end{bmatrix},
\qquad
Y_n^{-,\ev}
=
\begin{bmatrix}
\mathbf e_n\\
\kappa_n\mathbf e_n
\end{bmatrix}.
\label{eq:supp_free_ev_channels}
\end{equation}
Here \(Y_n^{+,\ev}\) decays as \(x\to+\infty\), while \(Y_n^{-,\ev}\) decays as \(x\to-\infty\).
Overall normalization factors are irrelevant for the rank argument.

For clarity, take a finite frequency cutoff.
Let \(N_B^{\free}(\epsilon)\) be the number of propagating sidebands in one lead, and \(N_E^{\free}(\epsilon)\) the number of evanescent sidebands.
The full free-lead WFV space at fixed quasienergy has dimension
\begin{equation}
2N
=
2\bigl(N_B^{\free}+N_E^{\free}\bigr).
\label{eq:supp_full_wfv_dim}
\end{equation}
The cutoff is only used to make the dimension count explicit; the genericity statement is unchanged as the cutoff is enlarged.

\subsection{Bound-state matching condition}
\label{subsec:supp_bound_matching}

A true Floquet bound state must be normalizable in both leads.
Thus it cannot contain any propagating component in either lead.
It also cannot contain evanescent waves that grow toward spatial infinity.

On the left lead, which extends to \(x\to-\infty\), the only allowed free components are the evanescent waves that decay to the left:
\[
Y_n^{-,\ev}.
\]
Let \(B_L^-\) be the matrix whose columns are these left-decaying WFV basis vectors evaluated at the left boundary:
\begin{equation}
B_L^-
=
\bigl[
Y_{n_1}^{-,\ev},
Y_{n_2}^{-,\ev},
\dots,
Y_{n_{N_E^{\free}}}^{-,\ev}
\bigr].
\label{eq:supp_BL_minus}
\end{equation}
Therefore the most general left-normalizable boundary vector is
\begin{equation}
Y_L=B_L^- c_L,
\qquad
c_L\in\mathbb C^{N_E^{\free}}.
\label{eq:supp_left_allowed}
\end{equation}

On the right lead, which extends to \(x\to+\infty\), the only allowed free components are the evanescent waves that decay to the right:
\[
Y_n^{+,\ev}.
\]
Let \(B_R^+\) be the matrix whose columns are these right-decaying WFV basis vectors evaluated at the right boundary:
\begin{equation}
B_R^+
=
\bigl[
Y_{n_1}^{+,\ev},
Y_{n_2}^{+,\ev},
\dots,
Y_{n_{N_E^{\free}}}^{+,\ev}
\bigr].
\label{eq:supp_BR_plus}
\end{equation}
The most general right-normalizable boundary vector is then
\begin{equation}
Y_R=B_R^+ c_R,
\qquad
c_R\in\mathbb C^{N_E^{\free}}.
\label{eq:supp_right_allowed}
\end{equation}

A bound state exists only if a left-decaying boundary vector, after propagation through the driven region, lands inside the right-decaying subspace.
Using Eq.~\eqref{eq:supp_spatial_propagation}, this condition is
\begin{equation}
F(\epsilon)B_L^-c_L
=
B_R^+c_R
\label{eq:supp_bound_match_direct}
\end{equation}
for some nonzero pair \((c_L,c_R)\).
Equivalently,
\begin{equation}
\mathcal M_b(\epsilon)
\begin{bmatrix}
c_L\\
c_R
\end{bmatrix}
=0,
\qquad
\mathcal M_b(\epsilon)
=
\bigl[
F(\epsilon)B_L^- \mid -B_R^+
\bigr].
\label{eq:supp_bound_matching_matrix}
\end{equation}
Thus a true Floquet bound state exists if and only if the two subspaces
\[
F(\epsilon)\mathrm{span}(B_L^-)
\quad\text{and}\quad
\mathrm{span}(B_R^+)
\]
have a nontrivial intersection.

This is the desired WFV-space formulation.
It uses only the fact that a Floquet eigenstate is spatially determined by its boundary WFV, and that free-space WFVs split concretely into propagating waves and evanescent waves.

\subsection{Overdetermination in an open quasienergy sector}
\label{subsec:supp_overdetermination}

We now count dimensions.
The matrix \(\mathcal M_b(\epsilon)\) has
\[
2N_E^{\free}
\]
columns, because both \(c_L\) and \(c_R\) have \(N_E^{\free}\) components.
Its rows span the full free-lead WFV space of dimension
\[
2N
=
2\bigl(N_E^{\free}+N_B^{\free}\bigr).
\]
Therefore
\begin{equation}
\mathcal M_b(\epsilon)
\in
\mathbb C^{
2(N_E^{\free}+N_B^{\free})
\times
2N_E^{\free}
}.
\label{eq:supp_Mb_size}
\end{equation}

If the quasienergy lies below all propagating continua, then \(N_B^{\free}=0\).
In that case \(\mathcal M_b\) is square, and the bound-state condition is the usual determinant condition
\[
\det \mathcal M_b(\epsilon)=0.
\]
Tuning the spectral parameter \(\epsilon\) can then generically produce isolated bound-state energies, as in an ordinary static bound-state problem.

The situation is different in an open quasienergy sector.
When at least one propagating sideband is present, \(N_B^{\free}>0\), and \(\mathcal M_b\) is a tall matrix.
A nonzero solution of Eq.~\eqref{eq:supp_bound_matching_matrix} requires
\begin{equation}
\mathrm{rank}\,\mathcal M_b(\epsilon)
<
2N_E^{\free}.
\label{eq:supp_rank_drop_condition}
\end{equation}
Equivalently, all \(2N_E^{\free}\times2N_E^{\free}\) maximal minors of \(\mathcal M_b(\epsilon)\) must vanish simultaneously.
This is no longer a single scalar condition.
It is an overdetermined set of constraints: one is trying to make two \(N_E^{\free}\)-dimensional decaying subspaces intersect inside a larger
\(2(N_E^{\free}+N_B^{\free})\)-dimensional WFV space.

Generically, two \(N_E^{\free}\)-dimensional subspaces in a
\(2(N_E^{\free}+N_B^{\free})\)-dimensional space have zero intersection when \(N_B^{\free}>0\).
Equivalently, the concatenated matrix
\(\mathcal M_b\) has full column rank for generic parameter values.
A nontrivial kernel appears only at exceptional points where the maximal minors vanish simultaneously. This typically cannot be achieved by tuning a single parameter $\epsilon$ alone, especially when the cutoff on the number of propagating sidebands $N_B^{\free}$ is taken arbitrarily large -- the number of independent algebraic constraints grows with the cutoff while only one parameter ($\epsilon$) is available to satisfy them. 
Unless a symmetry enforces these minors to vanish identically, such points form a lower-dimensional subset of the parameter space.

This is the precise sense in which true Floquet bound states in an open quasienergy sector are nongeneric.
They are Floquet analogues of bound states in the continuum: possible in specially engineered systems, but not stable under generic changes of boundary profile, driving amplitude, or sample parameters
\cite{Yajima1983,Kaneta1987,Della2014}.

Consequently, for a propagating branch in a generic open Floquet geometry, the bound-sector weight of a deep-bulk packet vanishes:
\begin{equation}
\mathcal P_\mu^{\rm b}(\epsilon_0)=0.
\label{eq:supp_no_bound_weight}
\end{equation}
Since the packet is normalized,
\begin{equation}
\mathcal P_\mu^{\esc}(\epsilon_0)
=
1-\mathcal P_\mu^{\rm b}(\epsilon_0)
=
1.
\label{eq:supp_escape_unity}
\end{equation}
Combining this with the deep-bulk branch-population principle
\[
p_\mu(\epsilon_0)=\mathcal P_\mu^{\esc}(\epsilon_0)
\]
gives the generic result
\begin{equation}
p_\mu(\epsilon_0)=1.
\label{eq:supp_generic_pmu_unity}
\end{equation}
This is the generic-openness statement used in the main text.

\section{Numerical check of generic openness}
\label{sec:supp_numerical_openness}

We finally give a numerical check of the generic-openness statement
\[
p_\mu(\epsilon_0)=1.
\]
The purpose of this section is not to prove generic openness, which follows from the overdetermined bound-state matching argument in Sec.~\ref{sec:supp_generic_openness}, but to show that the branch-population data exhibit the expected convergence toward unity in a representative open Floquet geometry.

We consider a sharply contacted driven lattice with bulk potential
\begin{equation}
V(x,t)
=
V\cos(2\pi x-\omega t)+V_0,
\qquad
V=8,\quad \omega=1,
\label{eq:supp_mc_model}
\end{equation}
and three offsets
\[
V_0=-1,\,-2,\,-3.
\]
Negative \(V_0\) shifts the frequency centers of some Floquet branches below the free-space propagation threshold. 
In a static problem, such below-threshold branches would be inaccessible from scattering states and would also be unable to escape into the leads.
In the Floquet problem, however, sideband coupling can connect these branches to open scattering channels.
The numerical question is therefore whether their total branch population still converges to unity.

The long-sample branch populations are evaluated by Monte Carlo sampling of the independent propagation phases that remain after shrinking-window smoothing \cite{Bjoint}.
For each Monte Carlo sample number \(N_{\rm MC}\), let
\[
\widehat p_\mu^{(N_{\rm MC})}
\]
denote the running estimator of the total population on branch \(\mu\).
To test convergence of all branches simultaneously, we monitor the extremal deviations from unity,
\begin{equation}
\begin{split}
\Delta_{\max}(N_{\rm MC})
&=
\left|
1-\max_\mu
\widehat p_\mu^{(N_{\rm MC})}
\right|,
\\
\Delta_{\min}(N_{\rm MC})
&=
\left|
1-\min_\mu
\widehat p_\mu^{(N_{\rm MC})}
\right|.
\end{split}
\label{eq:supp_mc_extremal_deviation}
\end{equation}
If some branch population converged to a value different from unity, then the corresponding quantity
\(\log |1-\widehat p_\mu^{(N_{\rm MC})}|\)
would eventually saturate to a plateau.
By contrast, for an unbiased Monte Carlo estimator converging to \(p_\mu=1\), the residual statistical error is expected to decrease as
\begin{equation}
\Delta(N_{\rm MC})\sim N_{\rm MC}^{-1/2},
\label{eq:supp_mc_scaling}
\end{equation}
up to a prefactor set by the variance of the sampled phase-torus function.

\begin{figure}[t]
    \centering
    \includegraphics[width=\linewidth]{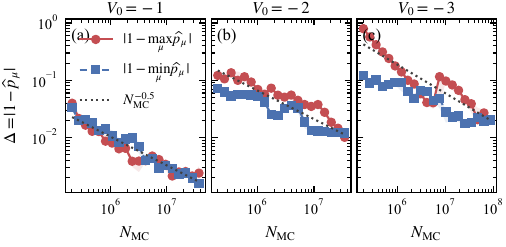}
    \caption{
    \textbf{Monte Carlo convergence of branch populations toward generic openness.}
    Log--log plot of the extremal deviations
    \(\Delta_{\max}=|1-\max_\mu\widehat p_\mu^{(N_{\rm MC})}|\)
    and
    \(\Delta_{\min}=|1-\min_\mu\widehat p_\mu^{(N_{\rm MC})}|\)
    as functions of the Monte Carlo sample number \(N_{\rm MC}\).
    The data are computed for the sharply contacted Floquet lattice
    \(V(x,t)=V\cos(2\pi x-\omega t)+V_0\), with
    \(V=8\), \(\omega=1\), and \(V_0=-1,-2,-3\).
    The dotted line is the reference Monte Carlo scaling
    \(N_{\rm MC}^{-1/2}\).
    No plateau is observed up to \(N_{\rm MC}=10^8\).
    Since a plateau in \(\log|1-p_\mu|\) would signal convergence to a value different from unity, the continued \(N_{\rm MC}^{-1/2}\)-type decay supports the generic-openness result \(p_\mu=1\).
    }
    \label{fig:supp_pmu_mc}
\end{figure}

Figure~\ref{fig:supp_pmu_mc} shows the log--log convergence of
\(\Delta_{\max}\) and \(\Delta_{\min}\) for the three offsets.
Both extremal deviations continue to decrease with the reference
\(N_{\rm MC}^{-1/2}\) scaling, and no saturation plateau is observed up to
\(N_{\rm MC}=10^8\).
This behavior supports the conclusion that both the largest and smallest branch-population estimates converge toward unity.
The convergence becomes slower and more irregular for more negative \(V_0\), which is consistent with increasingly narrow resonant structures on the phase torus; these resonances increase the effective Monte Carlo variance but do not produce a nonzero limiting deviation.

\end{document}